# Conducting interfaces between amorphous oxide layers and SrTiO$_3$(110) and SrTiO$_3$(111)

Mateusz Scigaj [a,b], Jaume Gázquez [a], María Varela [c,d], Josep Fontcuberta [a], Gervasi Herranz [a], Florencio Sánchez [a,*]

[a] *Institut de Ciència de Materials de Barcelona (ICMAB-CSIC), Campus UAB, 08193 Bellaterra, Spain.*

[b] *Dep. de Física, Universitat Autònoma de Barcelona, Campus UAB, 08193 Bellaterra, Spain.*

[c] *Materials Science and Technology Division, Oak Ridge National Laboratory, Oak Ridge, TN 37831, USA.*

[d] *Dep. Física Aplicada III & Instituto Pluridisciplinar, Universidad Complutense de Madrid, Madrid, 28040 Spain.*

* E-mail address: fsanchez@icmab.es

**ABSTRACT**

Interfaces between (110) and (111)SrTiO$_3$ (STO) single crystalline substrates and amorphous oxide layers, LaAlO$_3$ (a-LAO), Y:ZrO$_2$ (a-YSZ), and SrTiO$_3$ (a-STO) become conducting above a critical thickness $t_c$. Here we show that t$_c$ for a-LAO is not depending on the substrate orientation, i.e. t$_c$ (a-LAO/(110)STO) ≈ t$_c$(a-LAO/(111)STO) interfaces, whereas it strongly depends on the composition of the amorphous oxide: t$_c$(a-LAO/(110)STO) < t$_c$(a-YSZ/(110)STO) < t$_c$(a-STO/(110)STO). It is concluded that the formation of oxygen vacancies in amorphous-type interfaces is mainly determined by the oxygen affinity of the deposited metal ions, rather than orientational-dependent enthalpy vacancy formation and diffusion. Scanning transmission microscopy characterization of amorphous and crystalline LAO/STO(110) interfaces shows much higher amount of oxygen vacancies in the former, providing experimental evidence of the distinct mechanism of conduction in these interfaces.

## 1. Introduction

The discovery of a two dimensional electron gas (2DEG) at the interface between the (001)-oriented wide band gap insulators $LaAlO_3$ (LAO) and $SrTiO_3$ (STO) [1] has triggered a huge interest [2-8]. The polarity discontinuity at the LAO/STO(001) interface is the generally invoked scenario to explain the metallicity at the interface. Interestingly, 2DEGs have been also observed in (110)-oriented LAO/STO interfaces, which nominally do not present a polarity discontinuity [9-11], and also in (111)-oriented systems [9], thus opening new ways to modulate the interface properties by changing the crystal orientation [12]. In parallel, it has been found that conducting interfaces also form when amorphous oxides are deposited on STO(001) [13-18] and STO(110) [9] substrates, by pulsed laser deposition (PLD) or atomic layer deposition. Contrary to the epitaxial LAO/STO interfaces, the electrical conduction in the amorphous interfaces is likely originated by oxygen vacancies in the STO substrate, in the vicinity of the interface, caused by redox reactions during film growth [14-18]. Both amorphous and crystalline interfaces are found to be conducting when the capping layers have a thickness above a critical threshold $t_c$ [14-17]. However, the $t_c$ of amorphous interfaces on STO(001) depends on the oxygen pressure during deposition [16,17]. Here we have investigated the formation of amorphous interfaces on STO crystals with (110) and (111) orientation. While we find that $t_c$ changes with the chemical nature of the different amorphous oxides (LAO, STO and yttria stabilized zirconia – YSZ), it is insensitive to the crystalline orientation, either (110) or (111). At the same time, we used scanning transmission electron microscopy (STEM) in combination with electron energy loss spectroscopy (EELS) to gain more insights into the origin of the conductance at the amorphous oxide interfaces. Our experiments determined the oxygen vacancy profiles across the interfaces comprising STO(110) and amorphous and crystalline LAO. We thereby inferred the presence of a nanometric layer of oxygen vacancies at the STO interface, with much higher amount of vacancies in the STO(110) crystal capped with amorphous LAO than in crystalline LAO/STO(110) interfaces.

## 2. Experimental

A series of amorphous a-LAO, a-STO and a-YSZ (~7% $Y_2O_3$ molar) films of varying thickness (in the 1 - 8 nm range) were deposited at room temperature on STO(110) substrates by PLD (λ = 248 nm). In the case of a-LAO, the films were deposited simultaneously on STO(110) and STO(111). The single crystalline substrates underwent a thermal treatment in a dedicated furnace at 1100 °C for 2 h in air to obtain a morphology of steps and flat terraces [19]. In addition and prior to the deposition, the substrates were heated *in-situ* up to 500 °C at 0.5 mbar oxygen pressure to eliminate adsorbates, and subsequently cooled down overnight to room temperature at 200 mbar of oxygen. The films were deposited at room temperature; other growth conditions were kept identical to those used previously [9] for high quality crystalline LAO films on STO(110) and STO(111): $P_{O_2}$ = $10^{-4}$ mbar, 1 Hz laser repetition rate, substrate-target distance of 60 mm, and laser pulse energy and fluence of about 26 mJ and 1.5 $J/cm^2$, respectively. For all the deposited materials, LAO, STO and YSZ, *in-situ* reflection high energy electron diffraction (RHEED) showed patterns with a halo and without Bragg reflections, in agreement with their expected amorphous nature. Detailed microstructural characterization of an a-LAO/STO(110) sample was performed using a dedicated STEM, a Nion UltraSTEM, operated at 200 kV and equipped with $5^{th}$ order Nion aberration corrector and a Gatan Enfinium dual EEL spectrometer, which provides atomic-resolution Z-contrast imaging and EELS, allowing simultaneous real space studies of structure, chemistry and electronic properties. For comparison purposes, a crystalline LAO/STO(110) film grown using identical conditions as in Ref. [9] was also examined. Specimens for STEM observations were prepared by conventional thinning, grinding, dimpling and Ar ion milling. For transport measurements, the interfaces were contacted via ultrasonic wire bonding with Al wires, measuring the resistance by injection of current along STO[001] in (110)-interfaces, and along STO[11-2] in (111)-interfaces.

## 3. Results and discussion

The four series, corresponding to a-LAO/STO(110), a-STO/STO(110), a-YSZ/STO(110) and a-LAO/STO(111) samples, show metallic-like conduction above a critical thickness. The temperature-dependence of the sheet resistance $R_S$ of a representative conducting sample of each series is presented in Fig. 1. In all cases there is metallic-like behavior from room temperature to a few tens of Kelvin. A small resistance upturn, similar to those observed in crystalline interfaces [9], can be appreciated at around 10 K (Fig. 1b and 1c). The temperature dependence of the resistance, as well as its room-temperature values, are comparable to those reported for corresponding amorphous interfaces on STO(001) substrates [14,16,17].

Fig. 2a shows the thickness dependence of the room-temperature conductance of the interfaces between the different amorphous oxides and STO(110). The critical thickness $t_c$ is different for each type of interface. In the case of the a-LAO/STO(110) samples, the transition occurs at $t_c$ ≈1.5-2 nm (i.e., the a-LAO film with thickness t = 1.5 nm is insulating, whereas that with t = 2 nm is conducting). The transition occurs at larger thickness for a-YSZ: the samples with thickness t ≤ 2 nm are insulating whereas for t ≥ 3 nm are conducting. The critical thickness increases even more for a-STO (samples with t = 3 and 5 nm were insulating and conducting, respectively). In short, for amorphous layers on STO(11O) substrates we observe the following relationship: $t_c$(a-LAO) < $t_c$(a-YSZ) < $t_c$(a-STO). It is remarkable that the same correlation between critical thickness and materials has been reported for amorphous interfaces on

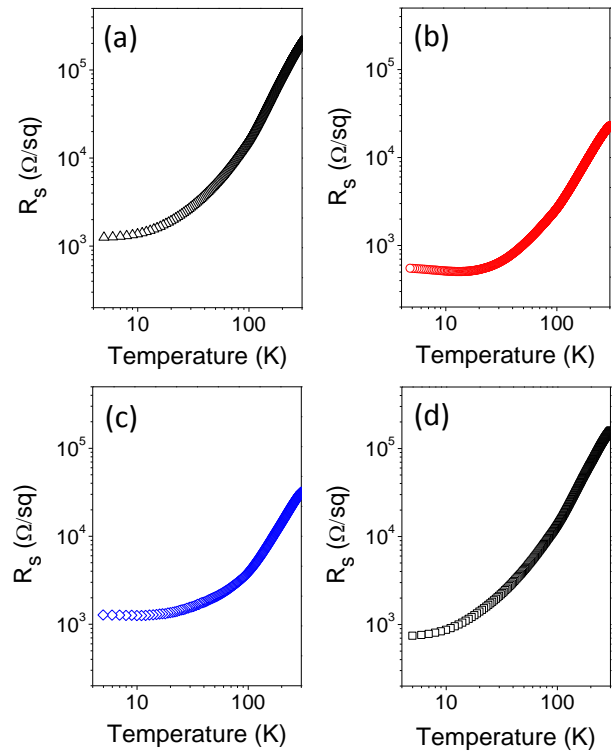

**Fig. 1.** Sheet resistance *versus* temperature for samples with 2 nm thick a-LAO (a), 7.8 nm thick a-STO (b) and 3 nm thick a-YSZ (c) layers on STO(110), and with 2 nm thick a-LAO (d) on STO(111).

STO(001), fabricated by PLD using fixed deposition parameters for all layers [14,16]. The redox reactions proposed as the mechanism for carrier generation in STO crystals capped with amorphous layers may depend on two factors: i) the oxygen affinity of the deposited atoms, and ii) the oxygen vacancy formation energy and diffusivity in the single crystalline substrate. Generally, vacancy formation and diffusion depend on the crystal orientation and therefore distinctive effects could occur at interfaces along different crystallographic orientations. However, the dependence of the conductance of a-LAO interfaces on both STO(110) and STO(111) substrates as a function of film thickness (Fig. 2b) indicates the same critical thickness for both orientations. Therefore, the combination of data shown in Fig. 2a and 2b clearly indicates that the capability to create oxygen vacancies and to induce the insulator-metal transition at the STO surface during growth of an amorphous layer is dictated by the oxygen affinity of the deposited cations, which dominates and overrules the distinct energy of oxygen vacancy formation and diffusion in each surface.

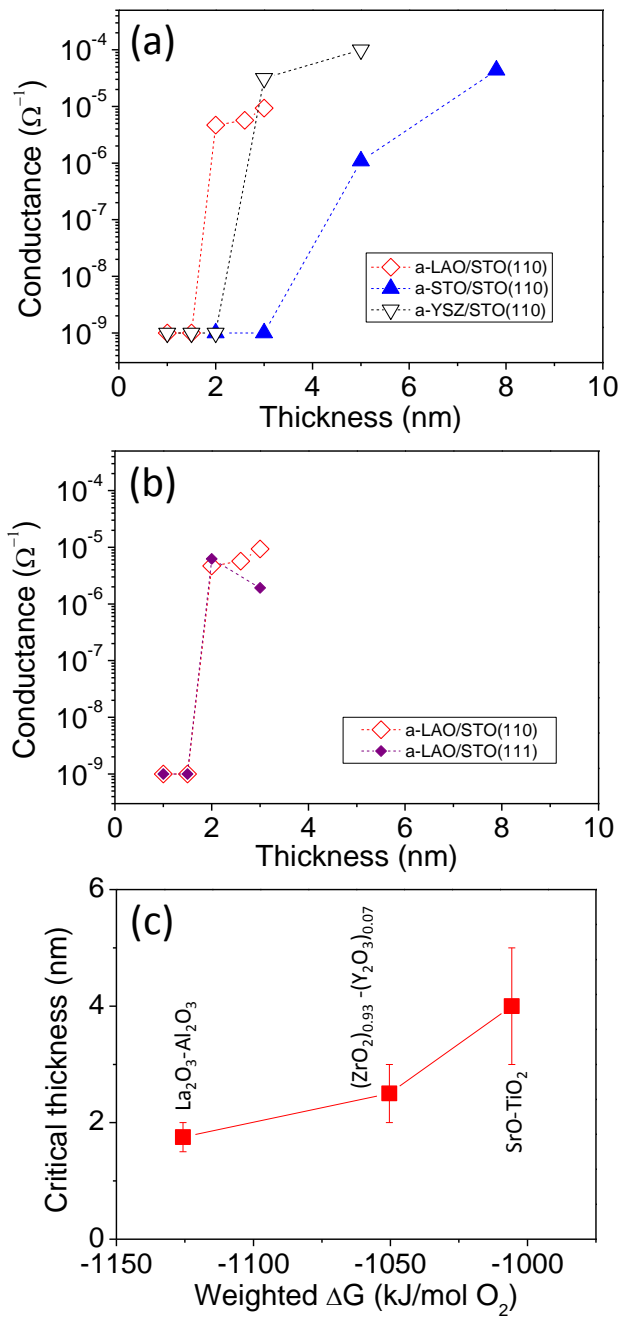

**Fig. 2.** (a) Room temperature conductance *versus* film thickness for samples with a-LAO (open rhombi), a-YSZ (open down triangles) and a-STO (solid up triangles) on STO(110). (b) Room temperature conductance *versus* thickness of a-LAO films on STO(110) (open rhombi) and STO(111) (solid rhombi). $10^{-9}$ $\Omega^{-1}$ is the measurement limit. (c) Critical thickness for a-LAO, a-YSZ and a-STO samples plotted against the Gibbs energy estimated from binary oxides of the present cations in each layer.

The change in the Gibbs energy for the oxidation reactions of each of the cations present in the amorphous layers has to be considered for a thermodynamic analysis. According to data [20] tabulated at 300 K the free energy of formation of $La_2O_3$, $Al_2O_3$, $ZrO_2$, $Y_2O_3$, SrO and $TiO_2$ are -1795.4 kJ/mol, -1581.7 kJ/mol, -1039.4 kJ/mol, -1816.1 kJ/mol, -561.2 kJ/mol and -889.1 kJ/mol, respectively. The formation of 1 mol of these compounds require different amount of oxygen [21]. Thus, we consider the free energy of formation of 2/3 mol of $La_2O_3$, $Al_2O_3$ and $Y_2O_3$, 1 mol of $ZrO_2$ and $TiO_2$, and 2 mol of SrO. On the other hand, the amorphous layers contain two metal ions and the two normalized Gibbs energies have to be averaged considering the relative content of each ion in each layer. The resulting weighted Gibbs energy of $La_2O_3$-$Al_2O_3$, $(ZrO_2)_{0.93}$-$(Y_2O_3)_{0.07}$ and SrO-$TiO_2$ are -1125.7 kJ/mol $O_2$, -1050.4 kJ/mol $O_2$ and -1005.7 kJ/mol $O_2$, respectively. The critical thickness corresponding to each deposited amorphous layer is plotted against the corresponding weighted Gibbs energy in Fig. 2c. The plot indicates that the critical thickness increases as the magnitude of the change in the Gibbs energy of oxide formation is lower, thus strongly supporting the view that critical thickness for electrical conduction is governed by the energy to create oxygen vacancies. At the same time, the fact that changes in the Gibbs energy are the most significant parameter for the creation of oxygen vacancies is perfectly coherent with the insensitivity of the critical thickness value with respect to crystal orientation.

Bearing in mind their pivotal role in the emergence of conductance, we exploited STEM-EELS to profile oxygen vacancies across the amorphous oxide interfaces. The low magnification STEM Z-contrast image of a nominally 4.8 nm thick a-LAO film on STO(110) shown in Fig. 3a indicates a well-defined interface between film and substrate and uniform film thickness of around 5 nm. The high resolution Z-contrast image in Figure 3b shows atomic columns along the [001] zone axis of the STO substrate, whereas no contrast can be appreciated within the LAO layer signaling absence of long range crystalline order. However, some atomic columns in the vicinity of the interface appear brighter than the rest in the STO substrate, signaling the presence of heavier La atoms, and indicating some crystalline order. STEM-EELS imaging reveals the chemistry of this interface. More specifically, panels c-e of Fig. 3 show atomic resolution maps of the La $M_{4,5}$, Sr $M_{4,5}$ and Ti $L_{2,3}$ edges, respectively. By overlaying all these mappings in Fig. 3f we can see how the different cations are distributed spatially, forming a crystalline La-rich layer at the interface. This La-rich layer is extremely thin, around 0.5 nm thick, well below the critical threshold for conducting interface in crystalline LAO/STO(110) [9] and, therefore, it cannot be the origin of the interface conductance.

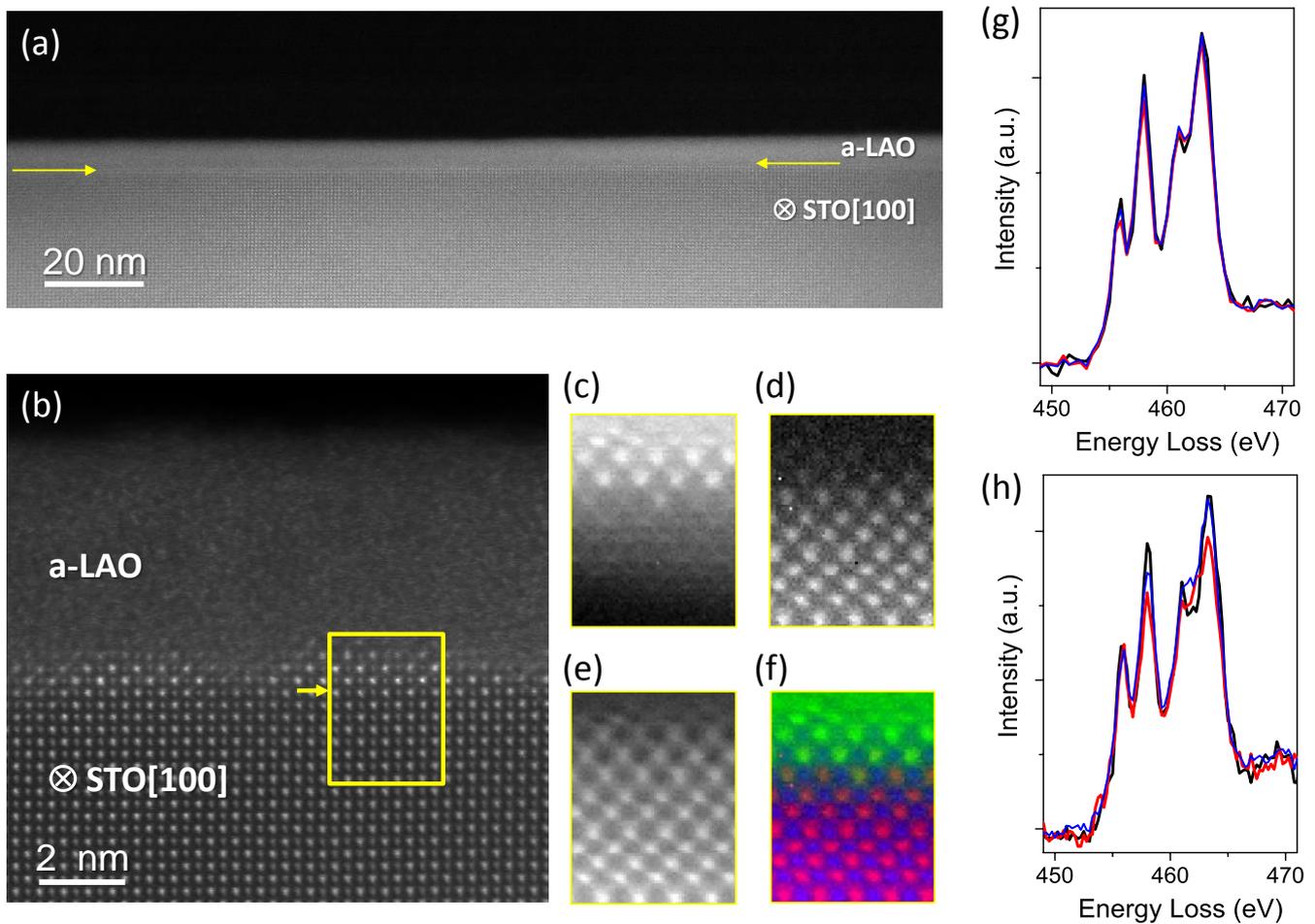

**Fig. 3.** STEM cross section view of a nominally 4.8 nm thick a-LAO/STO(110) along the STO[001] zone axis: (a) Low magnification Z-contrast image. Arrows point at the interface; (b) High resolution Z-contrast image of the a-LAO/STO interface. A square marks the area where the spectrum image was acquired, containing the interface. (c-e) Elemental maps corresponding to the La $M_{4,5}$, Sr $M_{4,5}$ and Ti $L_{2,3}$ edges, respectively. Integration windows 30 eV wide were used after background subtraction using a power-law fit. (f) RGB map produced by overlaying the Sr (in red), La (in green), and Ti (in blue). (g) and (h) show a set of Ti $L$ edge spectra from STO seven unit cells away from the interface (in black), two unit cells from the interface (in blue) and the Ti $L$ edge at the interface (in red) of epitaxial and amorphous grown LAO/STO(110) samples, respectively.

As stated above, oxygen vacancies generated by interface redox reactions are proposed for the observed conductance at (110) and (111) STO surfaces capped with amorphous layers. EEL spectroscopy allows extracting information about the presence of oxygen vacancies along the interface from the inspection of the shape of the Ti *L* edge, which reflects the underlying electronic structure. As each oxygen vacancy transfers two electrons to the Ti *d* band, the Ti *L* edge presents differences in the fine structure as Ti shifts from 4+ to a lower oxidation state [22]. Fig. 3g and h show a set of Ti *L* edge spectra from STO at different distances from the interface, seven unit cells away from the interface (in black), two unit cells from the interface (in blue) and the Ti *L* edge at the interface (in red) of epitaxial and amorphous grown LAO/STO(110) samples, respectively. Both figures show changes in the fine structure, notice the relative intensity between peaks, although the Ti *L* edge signal at the interface of the amorphous film presents greater differences when compared with the crystalline film. This is a clear hint on the presence of a higher $Ti^{3+}$ contribution at the amorphous interface, which is consistent with a much higher concentration of oxygen vacancies at the amorphous layer interface.

The presence of oxygen vacancies was also indirectly inferred by measuring the temperature-dependence of the conductivity of LAO on STO(110) samples under air atmosphere by heating the sample up to about 300 °C. It was found that a-LAO/STO(110) sample become insulating when heated above 250 °C, suggesting that annealing under oxidizing conditions removes oxygen vacancies in the amorphous interfaces and transforms irreversibly the interface from the conductive to the insulating state. Similar behavior was reported for a-LAO/STO(001) samples [14,17,18]. In sharp contrast, the conductivity of crystalline LAO/STO(001) was found to be stable and reversible in the same temperature range.

**Conclusions**

In conclusion, conducting interfaces are formed in amorphous oxide layers on STO(110) and STO(111) single crystalline substrates. We have observed here that the critical thickness does not depend on the substrate orientation but on the chemical nature of amorphous oxide. Moreover the critical thickness is found to decrease with the magnitude of the change in the Gibbs energy of oxide formation. This indicates that the oxygen affinity of the incoming ions, rather than the differences in vacancy formation and diffusion on a specific crystalline plane of the substrate, is the main factor determining the interface redox reactions that reduce the STO surface and trigger conduction at amorphous oxide/STO interfaces. The presence of oxygen vacancies in a nanometric layer of a STO(110) crystal capped with amorphous LAO has been experimentally confirmed by STEM-EELS.

**Acknowledgments**

Financial support by the Spanish Government [Projects MAT2011-29269-CO3 and MAT2014-56063-C2-1-R] and Generalitat de Catalunya (2014 SGR 734) is acknowledged. J.G. acknowledges the Ramon y Cajal program (RYC-2012-11709). Microscopy work has been conducted in the STEM Group of the Oak Ridge National Laboratory (ORNL). This work was supported by the U. S. Department of Energy, Office of Science, Basic Energy Sciences, Materials Science and Engineering Division (MV). Research at UCM is supported by the ERC Starting Investigator Award STEMOX 739239. We thank Dr. A. Pérez del Pino and Dr. R. Pfattner for collaboration in some transport measurements.